\documentclass{article}
\usepackage{spconf}
\usepackage{graphicx}
\usepackage{amsmath}
\usepackage{booktabs}
\usepackage{multirow}
\usepackage{tikz}
\usetikzlibrary{arrows.meta}
\usepackage{tabularx}
\usepackage{orcidlink}
\definecolor{codepurple}{RGB}{112,55,190}

\title{Towards Automated Clinical Behavioral Coding with Large Language Models: A Case study Using BOSCC recordings of Children}

\name{%
  \shortstack{%
    {\fontsize{10}{12}\selectfont
      Jordan Prescott$^{1}$\,\orcidlink{0009-0002-0914-4150},
      Aditya Kommineni$^{1}$\,\orcidlink{0009-0002-0056-8336},
      Tiantian Feng$^{1}$\,\orcidlink{0000-0002-2053-9068},
      Megan Micheletti$^{3}$\,\orcidlink{0000-0002-7197-7172},
      Alyssa Viggiano$^{2}$\,\orcidlink{0009-0004-2145-9395},}\\
    {\fontsize{10}{12}\selectfont
      \textit{%
        Luis Angeles$^{2}$\,\orcidlink{0009-0009-1038-2245},
        Anfeng Xu$^{1}$\,\orcidlink{0009-0001-5780-9797},
        Lynn Perry$^{2}$\,\orcidlink{0000-0001-6976-3741},
        Catherine Lord$^{3}$\,\orcidlink{0000-0001-5633-1253},
        Daniel Messinger$^{2}$\,\orcidlink{0000-0002-9551-675X},
        Shrikanth Narayanan$^{1}$\,\orcidlink{0000-0002-1052-6204}%
        \vspace{-3mm}
      }}%
  }%
}

\address{\fontsize{10}{12}\selectfont
    $^{1}$University of Southern California,\quad
    $^{2}$University of Miami,\quad
    $^{3}$University of California, Los Angeles
    \vspace{-1mm}
}

\begin{document}
\ninept

\maketitle

\begin{abstract}
% Autism Spectrum Disorder (ASD) is a neuro-developmental condition that leads to challenges with social communication, restricted interests and repetitive behaviors.
Autism spectrum disorder (ASD) is a neurodevelopmental condition characterized by differences in social communication and by restricted interests and repetitive behaviors.
Treatment interventions often target social-communication skills, creating a need for reliable measures of behavioral change. The Brief Observation of Social Communication Change (BOSCC) is a validated treatment-response measure based on brief play and social-communication interactions between a child and trained examiner.
% Typically this coding is performed manually by trained coders, which requires domain expertise and is resource-intensive.
% Hence, the ability to automate the scoring of these behavioral measures would facilitate the broader adoption and scalability of the BOSCC, thereby improving its accessibility and utility in both research and clinical settings.
The BOSCC coding process is resource-intensive and requires trained experts, motivating the automation of coding in order to improve scalability and accessibility.
% In this work, we evaluate general-purpose large language models (LLMs) for predicting speech-related BOSCC codes.
% Using transcripts and speaker roles of 163 in-house recordings, we find LLMs show promising potential towards automated scoring of behavioral codes.
In this work, we evaluate general-purpose large language models (LLMs) for predicting speech-related BOSCC codes from different input representations. We compare transcript, diarized-transcript, and targeted audio conditions across 163 in-house recordings.
% In this work, we investigate whether general-purpose large language models (LLMs) are able to automatically predict BOSCC codes relevant to speech-specific constructs from transcripts of BOSCC recordings.
% Across multiple LLM families evaluated on an in-house dataset of 163 BOSCC recordings, LLMs show promising potential towards automated scoring of behavioral codes.
LLMs are able to perform well in assessing verbal exchange, but do not perform as well when identifying atypical speech patterns. 
Additionally, performance varies considerably across scoring decisions, with no consistent pattern across diagnosis groups.
An audit of model predictions indicates that applying the BOSCC coding criteria and interpreting ambiguous speech evidence remain challenges.

\end{abstract}


\section{Introduction}
\label{sec:intro}

\begin{figure*}[h!]
    \centering
    \includegraphics[width=\linewidth]{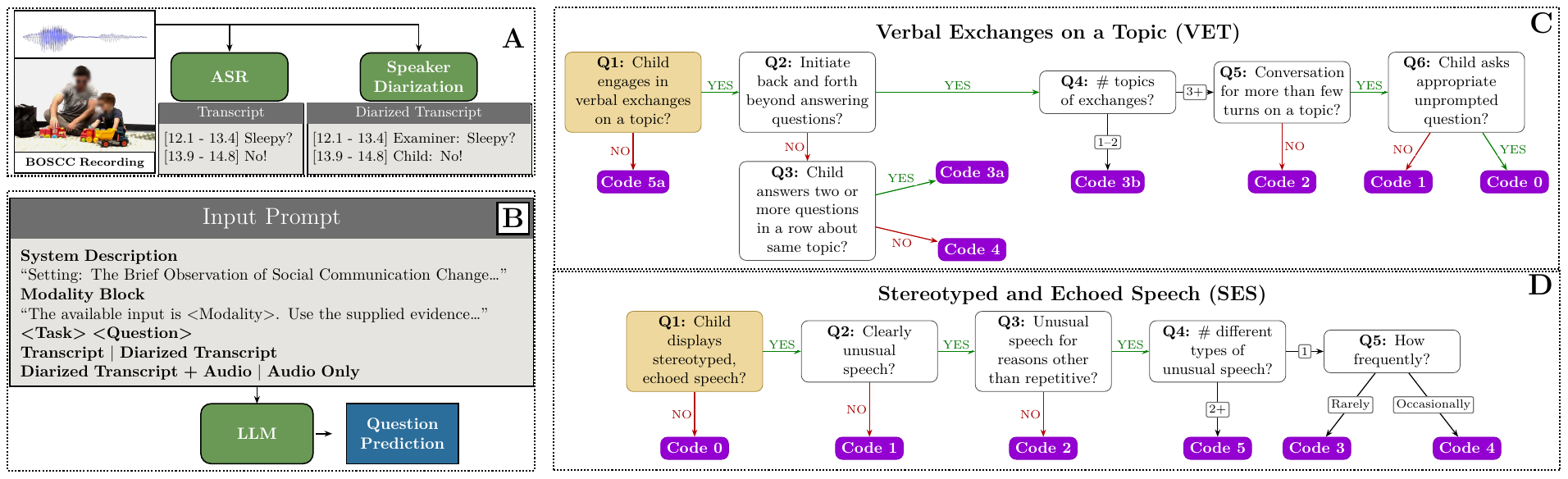}
    \caption{Overview of automated BOSCC item scoring pipeline. (A) ASR and speaker diarization produce transcript and diarized-transcript representations of each recording. (B) Each question-specific prompt presents the BOSCC task description, modality instructions, target decision question, and segment evidence, in that order. The LLM answers each question, and its linked responses determine a terminal code.     (C,D) Rubric decision trees showing the labeled question sequence and paths to terminal codes for Verbal Exchanges on a Topic (VET) and Stereotyped and Echoed Speech (SES), respectively.}
    \label{fig:schematic}
    % \vspace{-1mm}
\end{figure*}

Autism Spectrum Disorder (ASD) is a neuro-developmental disorder~\cite{lord2018autism} that is associated with difficulties in social communication, often characterized by repetitive behaviors (verbal and motor), and restricted interests~\cite{tedla2024assessing}.
Given the substantial impact of ASD on quality of life~\cite{lord2018autism, tedla2024assessing}, early identification and effective monitoring of behavioral changes over time are important.
To this end, Brief Observation of Social Communication Change (BOSCC)~\cite{boscc_orig,boscc_psyf} was proposed to track behavioral changes in social communication in autistic children over multiple time points during treatment.
% BOSCC is a validated measure for detecting change in autism symptoms during treatment.
It is administered by a clinician through a 12-minute play-based assessment and is video-recorded for subsequent scoring. 
Trained experts then score these videos on 20 distinct items that capture verbal and nonverbal communication, gaze behaviors, and restricted repetitive behaviors to characterize social communication patterns associated with ASD.

To score each of these items, experts follow behaviorally anchored rubrics to answer a sequence of logically connected questions, and the resulting decision path determines an ordinal score (0-5) [Figure~\ref{fig:schematic} (C,D) show the flowchart for \textit{Verbal Exchanges on a Topic} and \textit{Stereotyped and Echoed Speech} items], with a higher score indicating a greater severity of atypical behavior on that measure.
The specialized expertise and substantial time required for manual coding limit the ability to scale BOSCC administration~\cite{kim2019bosccados,toolan2025onlineboscc,jay2026boscc}, while the manual coding process may also be inefficient and is prone to human errors. 
With recent advances in deep learning and the development of foundation models for language, speech and vision~\cite{bommasani2021opportunities}, digitizing the BOSCC coding process and automating certain coding decisions would provide researchers with a more efficient and reliable approach for measuring changes in social communication.

Given the strong natural language understanding capabilities of Large Language Models (LLMs), recent work~\cite{aghakhani2025conversation, chen2026empirical, llm_clinical_scoring_todo,ali2025automatedcoding} has explored their potential for automated behavioral coding in healthcare applications.
Prior works in autism-specific studies have used LLMs for analyzing transcripts and multimodal recordings of child-adult interactions~\cite{feng2024genericllm,kommineni2024childinclusive,dosSantosMelicio2025ados,azizian2025multimodal,honarmand2026clinician}, 
with most studies focusing on autism-related diagnosis from video or the characterization of behaviors during child-clinician interactions.
While these studies highlight the potential of LLMs for automated behavioral analysis, their applicability to reliably support BOSCC coding remains unclear.

To address this knowledge gap, we evaluate whether general-purpose LLMs can reproduce expert-assigned BOSCC scores by answering the sequence of questions in each item's scoring rubric (refer to Figure~\ref{fig:schematic} (C,D)). 
% Focusing on two speech-related items, we find that zero-shot performance reaches up to 33.6\% balanced accuracy, an absolute improvement of 19.3 percentage points over chance.
% , although performance varies across models and items. % Results nevertheless vary substantially across models and constructs, and the evaluated interventions have mixed effects. 
% In particular, we show that speaker diarization can improve performance, whereas few-shot demonstrations or audio input do not provide consistent performance improvements.
% Question-level analysis reveals considerable variation across decisions, with model-provided evidence pointing to difficulties in applying the coding criteria and interpreting ambiguous inputs.
Our contributions are as follows:

\vspace{-1mm}
\begin{itemize}

    \item To assess LLMs for behavioral coding in child clinical settings, we evaluate their performance on two speech-related BOSCC items over 163 recordings. The best zero-shot balanced accuracy exceeds chance by 19.3 percentage points.

    \item We examine how ASR transcripts, speaker-role diarization, audio input, and few-shot prompting affect structured behavioral coding. Diarization improves 12 of 14 comparisons, while audio and few-shot prompting yield inconsistent gains.

    % \item We identify sources of error in LLM-based behavioral coding, including both overly strict or broad application of BOSCC criteria, and working from inaccurate ASR transcripts.

    \item We find LLM performance varies across scoring decisions, with no consistent diagnosis-group pattern. Analysis reveals difficulty in applying BOSCC criteria and ASR-related errors.

\end{itemize}

\section{Dataset}
\label{sec:data}
\vspace{-1mm}

% \begin{figure*}[h!]
%     \centering
%     \includegraphics[width=\linewidth]{img/Schematic_BOSCC.pdf}
%     \caption{Schematic of input prompt, model, and modified item tree structure.}
%     \label{fig:schematic}
% \end{figure*}

We used an in-house dataset of 163 BOSCC recordings collected using the Phrase Speech and Young Fluent (PSYF) version for analysis.
Children were grouped into typically developing (TD), autism spectrum disorder (ASD), and developmentally delayed (DD). DD and ASD are among those with Individualized Education Programs (IEPs) for special-education services~\cite{iep}. A documented ASD diagnosis determined assignment to ASD, the remainder were assigned to DD.
Table~\ref{tab:dataset_characteristics} summarizes the children's demographics and total duration of the dataset.
Each BOSCC recording lasted about 12 minutes and was divided into two six-minute segments that were coded independently.
Each segment contained semi-structured play and conversational interactions between the child and examiner~\cite{boscc_orig, boscc_psyf}.
In this study, we selected two items, \textit{Verbal Exchanges on a Topic} (VET) and \textit{Stereotyped and Echoed Speech} (SES) from BOSCC codes, that are relevant to assessing communication and speech-related behaviors. 

% \vspace{1mm}
\noindent \textbf{Verbal Exchanges on a Topic (VET)} assesses the child's participation in topic-related verbal exchanges with the examiner. A qualifying verbal exchange consists of a back-and-forth interaction between the examiner and child in which the child contributes at least two verbal turns associated with the topic. For example, the examiner might start with ``what's your favorite meal'' with the child responding, ``breakfast''. The examiner might then ask ``What do you like to eat for breakfast'' to which the child answers, ``pancakes''.

\vspace{1mm}
\noindent \textbf{Stereotyped and Echoed Speech (SES)} evaluates whether the child displays echoed, repetitive, or stereotyped verbal behavior during the exchange with the examiner. 

\vspace{1mm}
\noindent \textbf{Coding procedure.}
VET and SES are scored on a scale from 0 to 5. 
Scores were assigned by trained coders who completed consensus-based training and demonstrated the required level of agreement criterion on three consecutive videos before coding independently. 
Coders assign each score following an item-specific decision tree as shown for VET and SES in Figure~\ref{fig:schematic}(C, D).
% For each item, coders follow an item-specific decision tree. Figure~\ref{fig:schematic}(C,D) presents the trees for the two items examined in this study. 
Beginning at the root of the tree, the coder answers a sequence of questions about the child’s behavior and interactions during the six-minute segment. Each response directs the coder either to a subsequent question or to a terminal node which concludes the segmental-level score.

\begin{table}[t]
\centering
\footnotesize
% \scriptsize
\setlength{\tabcolsep}{5pt}
\vspace{-2mm}
\caption{BOSCC-PSYF corpus children demographics. Age is reported in years. Diagnosis counts are ordered as ASD/DD/TD.}
\label{tab:dataset_characteristics}
\begin{tabular}{@{}cccccc@{}}
\toprule
% \textbf{Ptp.}
\textbf{Obs.}
& \textbf{Segments}
& \textbf{Age}
& \textbf{Sex}
& \textbf{Diagnosis}
& \textbf{Hours} \\
\midrule
% 38
163
& 326
& 4.33$\pm$0.78
& 22M/16F
& 7/16/15
& 32.6 \\
\bottomrule
\end{tabular}
\vspace{-3mm}
\end{table}

\begin{table*}[t]
\centering
% \footnotesize
% \scriptsize
\setlength{\tabcolsep}{3pt}
\caption{Zero-shot terminal-code balanced accuracy (\%) on the full dataset using Parakeet ASR. Bracketed values are 95\% bootstrap percentile intervals from 10,000 resamples stratified by terminal code. Predictions use five-sample majority voting. Random-chance balanced accuracy is 14.3\% and 16.7\% for VET and SES, respectively. Best values are bolded and second-best values underlined.}
\label{tab:zero_shot_leaf_performance_parakeet_full}
\begin{tabular*}{\textwidth}{@{\extracolsep{\fill}}lcccc@{}}
\toprule
& \multicolumn{2}{c}{\textbf{Verbal Exchanges on a Topic}}
& \multicolumn{2}{c}{\textbf{Stereotyped and Echoed Speech}} \\
\cmidrule(lr){2-3}
\cmidrule(lr){4-5}
\textbf{Model}
& \textbf{Transcript}
& \textbf{Diarized Transcript}
& \textbf{Transcript}
& \textbf{Diarized Transcript} \\
\midrule

\textit{Random chance}
& \textit{14.3}
& \textit{14.3}
& \textit{16.7}
& \textit{16.7} \\
\midrule

% Qwen2.5-Omni-7B
% & $14.5\,[11.0,18.1]$
% & $15.3\,[11.1,19.8]$
% & $15.6\,[11.0,21.5]$
% & $16.1\,[10.7,21.8]$ \\

Qwen2.5-7B-Instruct
& $14.6\,[14.3,15.3]$
& $14.8\,[14.3,15.6]$
& $15.5\,[12.9,18.1]$
& $18.3\,[15.8,20.8]$ \\

Qwen3.5-9B
& $\underline{18.8}\,[14.7,23.3]$
& $30.8\,[25.9,36.1]$
& $\textbf{20.8}\,[13.2,28.8]$
& $18.7\,[11.2,26.6]$ \\

Qwen3.5-9B (thinking)
& $15.9\,[13.3,18.7]$
& $\textbf{33.6}\,[28.6,39.0]$
& $15.8\,[10.0,22.6]$
& $22.5\,[14.8,30.4]$ \\

Qwen3.8-27B-FP8
& $17.5\,[16.2,19.0]$
& $26.0\,[23.4,28.6]$
& $18.1\,[15.2,21.1]$
& $17.9\,[15.1,20.8]$ \\

Qwen3.8-27B-FP8 (thinking)
& $17.6\,[13.8,22.1]$
& $\underline{32.3}\,[28.6,35.9]$
& $17.1\,[14.3,20.0]$
& $22.4\,[18.6,26.6]$ \\

Gemma 4 12B-IT
& $\textbf{28.9}\,[22.3,35.6]$
& $\textbf{33.6}\,[28.9,38.6]$
& $\underline{20.6}\,[16.0,25.4]$
& $\underline{25.9}\,[18.9,33.4]$ \\

Gemma 4 12B-IT (thinking)
& $16.8\,[13.3,21.2]$
& $29.8\,[26.0,33.5]$
& $19.5\,[15.0,24.3]$
& $\textbf{26.3}\,[18.9,34.0]$ \\
\bottomrule
\end{tabular*}
% \vspace{-2mm}
\end{table*}

\vspace{1mm}
\section{Methods}

\subsection{BOSCC Decision Tree Scoring Formulation}

\noindent \textbf{Decision-tree formulation.}
To automate the BOSCC coding procedure for VET and SES, we use LLMs to reproduce the sequence of decisions that determine the final score of the item rather than directly predicting the final score.
% the decision process followed by BOSCC coders for the VET and SES items, we model the sequence of decisions leading to a score rather than the final score itself. 
Specifically, we decompose the decision trees shown in Figure~\ref{fig:schematic}(C,D) into binary questions, each with a question-specific prompt defining two possible responses. % for each binary question.
% To do so, we separate the decision trees shown in Figure~\ref{fig:schematic}(C,D) into their individual decision questions, each paired with a question-specific prompt that defines two possible responses. 
During inference, the LLM answers each question independently, without access to the responses or supporting evidence for earlier questions. 
For each segment, the resulting responses are linked according to tree structure to determine a terminal code and matching item score.
% the LLM’s responses to these questions can then be linked according to the tree structure to determine a terminal leaf.

\vspace{1mm}
\noindent \textbf{Decision-tree Structure and Evaluation.} Based on the BOSCC coding procedure, the VET decision tree contains six binary decision questions and seven terminal nodes. These include two distinct nodes corresponding to score~3 (3a and 3b). Meanwhile, the SES decision tree comprises five binary decision questions and six terminal nodes. We evaluate each question independently using all segments for which that question appears on the ground truth decision path. 
Specifically, a segment can be assigned to the correct terminal node only if the model answers every question along that path correctly.
% Reaching the correct terminal leaf requires a correct response to every linked decision question along its path.

\subsection{Modality Configurations}
\label{sec:mod}

We evaluated four speech-related input conditions: \textbf{transcript}, \textbf{diarized transcript}, \textbf{diarized transcript with audio}, and \textbf{audio-only}.
Figure~\ref{fig:schematic}(A) illustrates transcript and diarized transcript input construction using automatic speech recognition (ASR) and speaker-role diarization models, with ASR timestamps for both. 
For brevity, we refer to the latter as diarization throughout.
For all transcript-based conditions, audio recorded using a child-worn microphone was transcribed with NVIDIA Parakeet TDT 0.6B v2~\cite{parakeet}.

\vspace{1mm}
\par \noindent \textbf{Transcript} condition, each question-specific prompt included the Parakeet transcript for the complete six-minute interaction, consisting of ASR timestamped segments without speaker labels or other diarization information.

\vspace{1mm}
\par \noindent \textbf{Diarized transcript} condition concatenates timestamped speaker labels (child/examiner) generated by the child-adult diarization model of Xu et al.~\cite{Xu_2025} along with ASR inference. Utterance-level speech intelligibility labels from the ChildVox model~\cite{feng2026childvox}.
\vspace{0.5mm}
\par \noindent
Each resulting turn was represented as a compact JSON object.
When speaker labels overlapped for an ASR segment, both roles were reported in order of onset. 
Audio-quality flags for intelligible speech, unintelligible speech, and nonspeech were assigned, where multiple flags could apply to a turn.

\vspace{4mm}
\par \noindent
\textbf{Diarized transcript with audio} condition, we divided each six-minute recording into twelve consecutive, non-overlapping 30-second audio clips. 
The ASR and diarization outputs were partitioned using the time boundaries of each 30-second audio clip and paired with the corresponding audio clip in the input prompt.
\vspace{1.5mm}
\par \noindent \textbf{Audio-only} condition, the model processes the same audio clips as inputs without transcript information. All clips were converted to mono-channel and sampled at 16~kHz.

\subsection{Prompt Construction}

Our proposed prompt structure is shown in Figure~\ref{fig:schematic}(B).
Each question-specific prompt began with a shared BOSCC task preamble that introduced the problem setting, defined the child and examiner roles, and instructed the LLM to base its judgment only on provided input. 
This was followed by modality-specific instructions describing the available input and how it should be interpreted.
The prompt then presented the corresponding decision question, manually derived operational definitions and exclusions, coding guidance, the two permitted responses, and a fixed JSON output format. The LLM was instructed to return both a categorical response and reasoning. Only the categorical response was used for quantitative evaluation. Question wording, coding guidance, response labels, and output schemas were kept consistent across LLMs. 
% Segment evidence was provided last. 

For the transcript and diarized transcript conditions, we evaluated each full six-minute segment in a single prompt. For inference with audio modality, we performed inferences on each audio clip within a segment, which were then aggregated to produce a single response.
For zero-shot experiments, the prompts contained no labeled examples from the corpus. 
% When including demonstrations, example evidence and its corresponding answer were placed before the target evidence in balanced pairs containing one example for each possible response.
For prompts with demonstrations, balanced example pairs, one for each possible label, were presented before the ASR or diarization evidence.
% each decision question was first evaluated using twelve interval-level prompts, one for each 30-second clip. In the audio-augmented condition, each prompt contained an audio clip and its aligned ASR and diarization information, whereas the audio-only prompts contained only the clips. 
% The twelve LLM outputs were combined in a final prompt that produced a single response to the decision question.
% The twelve LLM outputs were then aggregated to produce a single response to the decision question.

% The accompanying repository provides the prompts, with dataset-specific content removed or modified to protect participant privacy.
% The modality instructions, supplied evidence, and evidence-processing procedure varied across input conditions. 

\section{Experimental Design}
\label{sec:design}

% \subsection{Models and Inference}
\par \noindent \textbf{Models and Inference.}
We evaluated three Qwen models, Qwen2.5-7B-Instruct~\cite{qwen2.5}, Qwen3.5-9B~\cite{qwen3.5}, and Qwen3.8-27B~\cite{qwen38}, along with the audio-capable Gemma 4 12B-IT model~\cite{team2026gemma}.
% We evaluated models from Qwen and Gemma model families, which include both text-only and omni-models.
% Evaluated text-only models include Qwen2.5-7B-Instruct, Qwen3.5-9B and Qwen3.8-27B, and omni-models include Gemma 4 12B-IT.
Inference for text-only models involved a cascaded pipeline with ASR and speaker labels generated by task-specific models as stated in Section~\ref{sec:mod}.
Corresponding transcripts were then input to LLMs for the prediction of different BOSCC codes.
Non-audio models were evaluated in transcript and diarized transcript conditions whereas Gemma 4 12B-IT was evaluated under all four modality conditions.
% Gemma 4 12B-IT was evaluated under three different modality settings, (1) text-only wherein the transcripts or diarized transcript are used as input, (2) audio-only where the audio recording is the only input provided and (3) text + audio wherein audio recordings and transcripts are both provided as inputs.
Additionally, in order to assess the effect of chain-of-thought thinking, models that support thinking mode (Qwen3.5-9B, Qwen3.8-27B and Gemma 4 12B-IT) were evaluated with or without thinking enabled.
We used the recommended sampling parameters from the original technical report or code repository.

% except Qwen2.5-7B-Instruct, which uses a temperature of 0.5 due to limited documentation.

vLLM (v0.28.0)~\cite{vllm} was used for model inference with Qwen3.8 loaded at FP-8 quantization and all other models at bfloat16.
Non-thinking generations were limited to 8,192 tokens, while thinking configurations allocated 8,192 tokens for reasoning and 512 for the final answer. 
For each segment-question pair, we repeated the inferences five times and used the majority label as the question-level prediction.
% extracted the permitted answer label using a fixed parser, 
% Inference used an 8,192-token budget across LLMs. For each segment-question pair, we sampled five responses and used the most frequent parsed answer label as the question-level prediction.
% Zero-shot and audio experiments used a 32,768-token context window, while the longer few-shot prompts used 49,152 tokens. Non-thinking outputs were limited to 8,192 new tokens. Thinking configurations received an 8,192-token reasoning budget followed by a 512-token reserve to ensure an answer and prevent vote truncation. We sampled five responses for every segment-question pair. A fixed parser extracted the permitted answer label from each response, and the most frequent label across the five samples was used as the question-level prediction.

% \subsection{Evaluation Protocol}
\par \noindent \textbf{Evaluation Protocol.}
% First, all models were evaluated under a zero-shot text-only setting under \textit{transcript} and \textit{diarized transcript} settings. 
Our primary experiment compared zero-shot performance across LLMs under the transcript and diarized transcript conditions using all 326 segments. After majority voting independently at each binary question, a terminal-code prediction was considered correct only when all question-level predictions along the reference path matched the corresponding reference answers.
Because the terminal-code distributions were imbalanced, we used balanced accuracy as the primary metric. We computed it as the unweighted mean recall across the seven VET codes and six SES codes.
We report 95\% bootstrap percentile confidence intervals from 10,000 bootstrap resamples stratified by ground truth terminal code.

Based on the comparison of zero-shot performance across LLMs, we conducted three analyses on few-shot prompting, modality effects, and prediction behavior using the strongest zero-shot LLM, Gemma 4 12B-IT. 
First, we evaluated two-shot and four-shot prompting with diarized transcripts. 
Second, we compared the diarized transcript condition with the audio-augmented and audio-only conditions. 
Finally, under the diarized-transcript condition, we examined exact terminal-code matching by diagnosis group, followed by question-level analysis of balanced accuracy and response patterns. We then manually inspected model-provided evidence and audio examples for errors on each item's worst-performing question.

\section{Results and Discussion}
\label{sec:results}

\subsection{Zero-shot Results}
Table~\ref{tab:zero_shot_leaf_performance_parakeet_full} reports zero-shot terminal-code balanced accuracy for VET and SES items under transcript and diarized transcript conditions across LLMs.
Results for both VET and SES indicate noticeable gains for the diarized transcript setting highlighting the importance of speaker-role information.
This improvement is more pronounced for VET, which achieves higher average performance than SES under diarization, with the two items perform similarly with transcripts alone.
We find the effect of thinking mode inconsistent, where it tends to help under diarized transcripts but providing no clear benefit with transcripts alone.
Gemma 4 12B-IT achieves the strongest average performance across the four settings, and is therefore used for the subsequent analyses.
In summary, these results indicate that current LLMs can capture meaningful BOSCC scoring signal, but their zero-shot performance across terminal codes remains insufficient for standalone automated BOSCC scoring.

\begin{table}[t]
\centering
% \footnotesize
% \scriptsize
\setlength{\tabcolsep}{5pt}
\caption{Terminal-code balanced accuracy (\%) of ablations for Gemma 4 12B-IT on the full
dataset. Parentheses report percentage-point changes from the zero-shot
diarized-transcript baseline.}
\label{tab:ablations}
\begin{tabular}{@{}lrr@{}}
\toprule
\textbf{Condition} & \textbf{VET} & \textbf{SES} \\
\midrule
Diarized transcript & 33.6 & 25.9 \\
% \addlinespace
\midrule
\multicolumn{3}{@{}l}{\textit{Demonstrations}} \\
\quad 2-shot & 33.5 ($-0.1$) & 24.9 ($-1.0$) \\
\quad 4-shot & 37.9 ($+4.3$) & 25.2 ($-0.7$) \\
\addlinespace
% \midrule
\multicolumn{3}{@{}l}{\textit{Audio input}} \\
\quad Diarized transcript + audio
    & 34.1 ($+0.5$) & 18.0 ($-7.9$) \\
\quad Audio only
    & 15.6 ($-18.0$) & 16.7 ($-9.2$) \\
\bottomrule
\end{tabular}
% \vspace{3mm}
\end{table}

\subsection{Ablations}
% \par \noindent \textbf{Ablations.}
Table~\ref{tab:ablations} summarizes the few-shot and audio-modality experiments with Gemma 4 12B-IT.
We observe that few-shot prompting does not provide consistent gains, with performance decreases in all settings except four-shot VET. Further, adding audio to diarized transcript yields only a small improvement for VET prediction and decreases performance on SES prediction, while audio-modality alone performs substantially worse than diarized transcripts for both items. Overall, Gemma 4 12B-IT does not show consistent benefit from either few-shot demonstrations or adding audio as input.

\subsection{Analysis of Scoring Decisions}
\begin{figure}
    \centering
    \includegraphics[width=\linewidth]{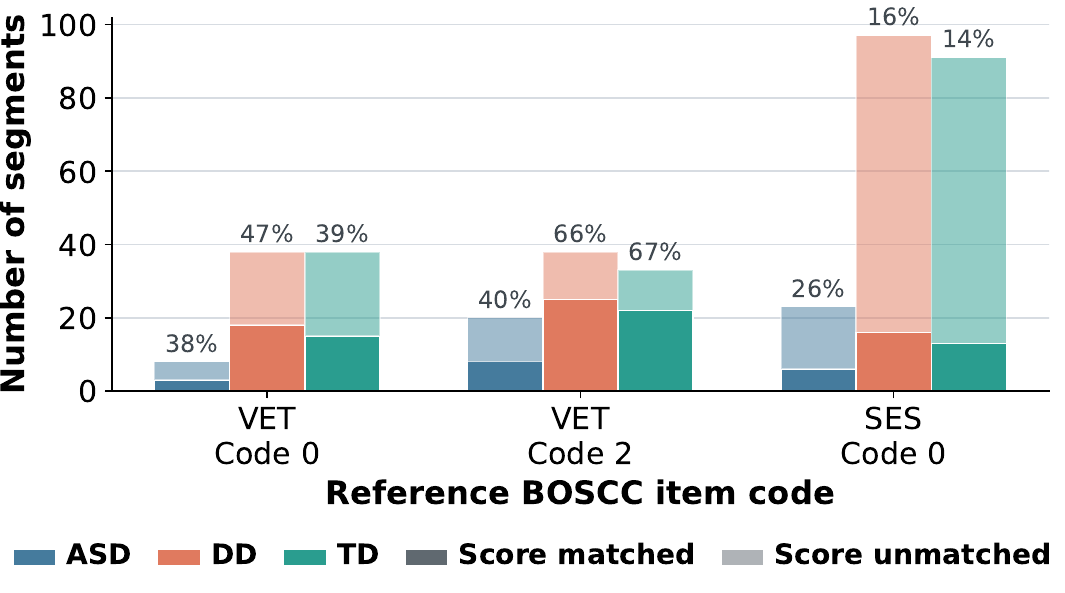}
    \vspace{-2mm}
    \caption{Exact BOSCC item-code matching by diagnosis for Gemma 4 12B-IT. Only BOSCC codes with at least 75 segments are shown. Labels report matched count over total group count in \%.}
    \label{fig:diag}
    % \vspace{-4mm}
\end{figure}

% \par \noindent \textbf{Analysis of Scoring Decisions.}
We focus on Gemma 4 12B-IT under the diarized-transcript condition. 
Figure~\ref{fig:diag} 
shows how well predicted codes match ground truth final scores by diagnosis on well supported terminal codes.
The performance for each code appears more affected by item and code rather than by diagnosis group, with no consistent trend on how diagnosis groups perform relative to each other.

To identify which binary decision questions contribute most to errors in final-score predictions, we report balanced accuracy for each decision question in Table~\ref{tab:gemma_question_performance}. 
The comparisons show that prediction performance varies substantially across questions, with Q3 yielding the lowest balanced accuracy for both VET and SES.
% the worst for each item by balanced accuracy being VET Q3 and SES Q3. 
Specifically, Q3 in VET asks whether the child answers at least two linked questions on the same topic. The model predicts \textsc{no} in 21 of 22 cases and fails to identify any reference \textsc{yes} cases. 
Meanwhile, Q3 in SES asks whether the child exhibits an unusual speech type besides repetition, such as echoed or stereotyped speech. In contrast, the model overpredicts \textsc{yes} and misclassifies 21 of the 28 reference \textsc{no} cases, leading to below-chance balanced accuracy.
% reports accuracy and balanced accuracy for each decision question. 

Manual inspection of model-provided rationale and input evidence revealed different limitations.
For VET Q3, the LLM's application of a BOSCC-defined back-and-forth exchange was inconsistent, sometimes requiring a more explicit relationship between successive questions than the criteria specify. 
For example, the examiner asked a child which door to open, the child answers ''number one,'' and the examiner followed by asking, ''Where is number one?'' The model does not treat this follow-up question as the same exchange and therefore answered incorrectly.
For SES Q3, the LLM applied the question criterion too broadly, treating ordinary task-related speech as evidence of atypical language. 
% Moreover, errors in predictions involve incomplete, ambiguous, or mistranscribed utterances,
For both questions, listening to audio examples exposed errors associated with ambiguous or inaccurate ASR output.
For example, the phrase “A cookie!” was transcribed as “I’m cookie”, which the LLM treated as unusual self-reference.
These examples show how scoring errors can arise from both the model's interpretation and errors in the given evidence.

\begin{table}[t]
\centering
% \footnotesize
\setlength{\tabcolsep}{4pt}
\caption{Question-level balanced accuracy (BA; \%) for Gemma 4 12B-IT under
the diarized-transcript condition. $n$ is the number of segments whose
reference path reaches each question.}
\label{tab:gemma_question_performance}
\begin{tabular*}{\columnwidth}{@{\extracolsep{\fill}}lcclcc@{}}
\toprule
\multicolumn{3}{c}{\textbf{VET}}
& \multicolumn{3}{c}{\textbf{SES}} \\
\cmidrule(lr){1-3}
\cmidrule(lr){4-6}
\textbf{Question} & \textbf{$n$} & \textbf{BA}
& \textbf{Question} & \textbf{$n$} & \textbf{BA} \\
\midrule
Q1 & 326 & 84.2 & Q1 & 326 & 55.3 \\
Q2 & 303 & 50.5 & Q2 & 115 & 57.2 \\
Q3 &  22 & 44.4 & Q3 &  69 & 49.1 \\
Q4 & 281 & 52.0 & Q4 &  41 & 54.2 \\
Q5 & 219 & 66.7 & Q5 & 34 & 73.5 \\
Q6 & 128 & 66.7 & \hspace{0.35mm} -- & -- & -- \\
\bottomrule
\end{tabular*}
\end{table}

\section{Conclusion}

We evaluated whether LLMs can score two speech-related BOSCC items by answering their linked decision questions. 
% We evaluated LLM scoring of two speech-related BOSCC items through linked decision questions.
Qwen and Gemma models captured meaningful zero-shot signal but remained insufficient for standalone use. 
Diarized transcripts provided the most consistent gains, while demonstrations and audio yielded no consistent benefit across items.
Among the well-supported codes, performance showed no consistent pattern by diagnosis group. 
Inspection of LLM evidence revealed inconsistent criterion application and sensitivity to ambiguous speech evidence. 
These findings are limited to two items, one in-house dataset, and single-coder annotations.
Future work should explore models with improved acoustic integration and improved ASR, extending evaluation to additional BOSCC items and datasets.

\newpage

\section{Compliance with Ethical Standards}
This study used previously collected BOSCC recordings from child participants. Data collection and analysis were approved under IRB Study No.~20160509. Written informed consent was obtained from each participant’s parent or legal guardian. Recordings were accessed only by authorized study personnel and analyzed in accordance with the approved protocol.

\section{Acknowledgments}
This work was supported by the National Science Foundation Graduate Research Fellowship Program grant DGE-1842487, the Simons Foundation Autism Research Initiative under award SFI-AR-HUMAN-00004115-01, the National Science Foundation under award 2150830, and the Eunice Kennedy Shriver National Institute of Child Health and Human Development under award R01HD057284.

\bibliographystyle{IEEEbib}
\bibliography{refs}

\end{document}